\documentclass[%
reprint,
superscriptaddress,
amsmath,amssymb,
prl,
floatfix,
]{revtex4-2}

\usepackage{graphicx}
\usepackage{latexsym}
\usepackage{amsmath}
\usepackage{graphics}
\usepackage{amssymb}
\usepackage{layout}
\usepackage{verbatim}
\usepackage{amsfonts,epsfig,dsfont,bbm}
\usepackage{natbib}
\usepackage{hyperref}
\usepackage{color}
\usepackage[dvipsnames]{xcolor}
\usepackage{siunitx}
\usepackage{diagbox}
\usepackage{cleveref}
\usepackage[normalem]{ulem}

\def\tred{\textcolor{red}}

\begin{document}


\title{
Data-driven criteria for quantum correlations}

\author{Mateusz Krawczyk}
\affiliation{Institute of Theoretical Physics, 
Wroc{\l}aw University of Science and Technology,
50-370 Wroc{\l}aw, Poland}

\author{Jarosław Paw\l{}owski}
\affiliation{Institute of Theoretical Physics, 
Wroc{\l}aw University of Science and Technology,
50-370 Wroc{\l}aw, Poland}

\author{Maciej M. Ma\'{s}ka}
\affiliation{Institute of Theoretical Physics, 
Wroc{\l}aw University of Science and Technology,
50-370 Wroc{\l}aw, Poland}

\author{Katarzyna Roszak}
\affiliation{Institute of Physics (FZU), Czech Academy of Sciences, Na Slovance 2, 182 00 Prague, Czech Republic}

\date{\today}

\begin{abstract}
We build a machine learning model to detect correlations
in a three-qubit system using a neural network trained in an unsupervised manner on randomly generated states.
The network is forced to recognize separable states, and correlated states are detected as anomalies.
Quite surprisingly, we find that the proposed detector performs much better at distinguishing a weaker form of quantum correlations,
namely, the quantum discord, than entanglement. In fact, it has a tendency to grossly overestimate the set of entangled states
even at the optimal threshold for entanglement detection, while it underestimates the set of discordant states to a much lesser
extent. 
In order to illustrate the nature of states classified as quantum-correlated, 
we construct a diagram containing various types of states -- entangled, as well as separable, both discordant and non-discordant.
We find that the near-zero value of the
recognition loss reproduces the shape of the non-discordant separable
states with high accuracy, especially considering the non-trivial shape of this set on the diagram. 
The network architecture is designed carefully: it preserves
separability, and its
output is equivariant with respect to qubit permutations.
We show that the choice of architecture is important to get
the highest detection accuracy, much better than for a baseline model that just utilizes a partial trace operation.
\end{abstract}

\maketitle
Quantum correlations play a pivotal role in near-future quantum technologies, from quantum cryptography \cite{gisin02,pirandola20}, networks \cite{kimble08,wehner18}, to computation \cite{nielsen00,demirel21}.
The source of their usefulness is that they allow for the state of one subsystem to be influenced by operations and measurements
on the other, which yields effects that cannot be reproduced by classical physics. A good example here is teleportation \cite{bennett93,bouwmeester97,boschi98},
which allows for the transfer of an unknown quantum state between distant locations by means of local operations and
measurements, and the transfer of classical information, due to a maximally entangled state being shared by two parties. 

In quantum information theory, there is a long-standing question about quantum correlations in mixed states \cite{horodecki05,modi12}.
For pure states, any correlations between subsystems are quantum in their nature (entanglement), and
the set of separable (not entangled) states is equivalent to the set of product states. 
In the case of mixed states, the fact that a density matrix cannot be written in product form does not
guarantee the existence of quantum correlations, since this formalism allows for the description of classical (purely statistical) correlations, i.e.~$\hat{\rho}_{AB}=1/2\left(|00\rangle\langle 00|+|11\rangle\langle 11|\right)$,
whereas a product state contains no correlations whatsoever. 

One definition of the set of states that contain no quantum correlations, namely the set of mixed separable states, is 
given by \cite{werner89,plenio07,horodecki09}
\begin{equation}
    \label{sep}
    \hat{\rho}_{AB}^{sep}=\sum_ip_i|\psi_i\rangle_{AA}\langle\psi_i|\otimes|\phi_i\rangle_{BB}\langle\phi_i|,
\end{equation}
where $p_i$ are probabilities and the indices $A$ and $B$ label the subsystems. 
It is relevant to note here that there is no limitation on states $|\psi_i\rangle_{A}$ or $|\phi_i\rangle_{B}$,
which do not have to form a basis. By this definition, a bipartite state is separable if and only if it can be created
from a product state using only local operations and classical communication. 
The quantification of entanglement (for mixed states, entangled states are defined as not-separable) is classed as an NP-complete problem \cite{gurvits04}, and it does not simplify even when a state only needs to be qualified as separable or entangled.

A state with no quantum correlations can also be defined as a state for which no information about one subsystem can
be learned by measurements performed on the other subsystem. Such states are said to contain zero \textit{discord} \cite{ollivier01,henderson01,modi14}, and
are defined by Eq.~(\ref{sep}) with the added constraints that the states $|\psi_i\rangle_{A}$ must be part of
an orthonormal basis, and the same for states $|\phi_i\rangle_{B}$. Quantifying the amount of discord in a given state is also NP-complete \cite{huang14}, but it
is harder than quantifying entanglement for small systems \cite{dakic10,miranowicz12,tufarelli13}. On the other hand, simply qualifying a state as discordant or not is a straightforward task \cite{dakic10,huang11}. The set of non-discordant states is a subset of separable states.
Currently, entanglement is the accepted, standard metric for quantum correlations, yet it should be noted that
discord has one important advantage over entanglement, namely that it has a much stronger relationship to measurable quantities,
while the detection of entanglement generally requires quantum tomography \cite{james01,christandl12}.

In the following, we would like to go beyond the typical scheme of describing the correlations in quantum states using analytical metrics and propose the study of quantum correlations using machine learning (ML) models forcing them to learn similarities between different classes of quantum states.
We use neural networks (NNs) which have proven their usefulness in many applications, such as computer vision or signal processing~\cite{Chai2021,Alam2020}, but recently also in quantum physics, e.g., in many-body theory~\cite{Carleo2017,Gao2017,Levine2022, cheng2023} or in entanglement modelling~\cite{Levine2019,Chen2022,siamese2022}. However, in our approach, we carefully design the model architecture to be separability-preserving and qubit permutations equivariant, which is part of the general trend of building NN models that respect symmetries of a given problem, for example, being
rotation and translation group-equivariant \cite{cohen2016,bekkers2018,cohen2019,nagai2023}.

We present results where
the ML model is first trained to distinguish between separable and entangled states (on both mixed and pure states)
and then tested on similar sets. These sets (both in the training and testing phases) apart from entangled states, contain different variants of separable states, with special emphasis lain on discordant
separable states. 
While the ML model trained only on pure states can only distinguish between
product and non-product mixed states, hence signifying correlations, but not being
able to discriminate between classical, quantum-classical (discord), and
quantum-quantum (entanglement) correlations,
quite surprisingly, we find that the performance of the separator
trained also on mixed states
is much better at signifying quantum discord than
entanglement.
This corroborates the intuition that a type of quantum correlations which are defined on the properties of measurement outcomes, as opposed to correlations defined on specifics of state preparation, should be more evident in a given state, even though the ML model was not trained directly on non-discordant states (but on separable ones).

Since the goal of using the neural network is not only to classify the discordance or entanglement of states, but also to produce a similarity measure, we propose to train the model in the unsupervised anomaly detector scheme~\cite{Pankaj2015}. Anomaly detection is a standard problem in ML where a model is trained to reconstruct normal data and then fails when trying to reconstruct abnormal data (anomaly). The reconstruction error, i.e. the model ``loss function'' measures the similarity of a given unknown sample to the training data, and if it is greater than some assumed threshold, then the sample is treated as an anomaly. In our case, anomalies are states that are quantum-correlated.
The model is trained to recognize separable states of three qubits. 
We study 3-qubit states to have a non-trivial yet manageable Hilbert space, keeping in mind that the proposed model construction
can be easily extended to larger qubit registers. 

\begin{figure}[!bt]
    \centering
\includegraphics[width=\columnwidth]{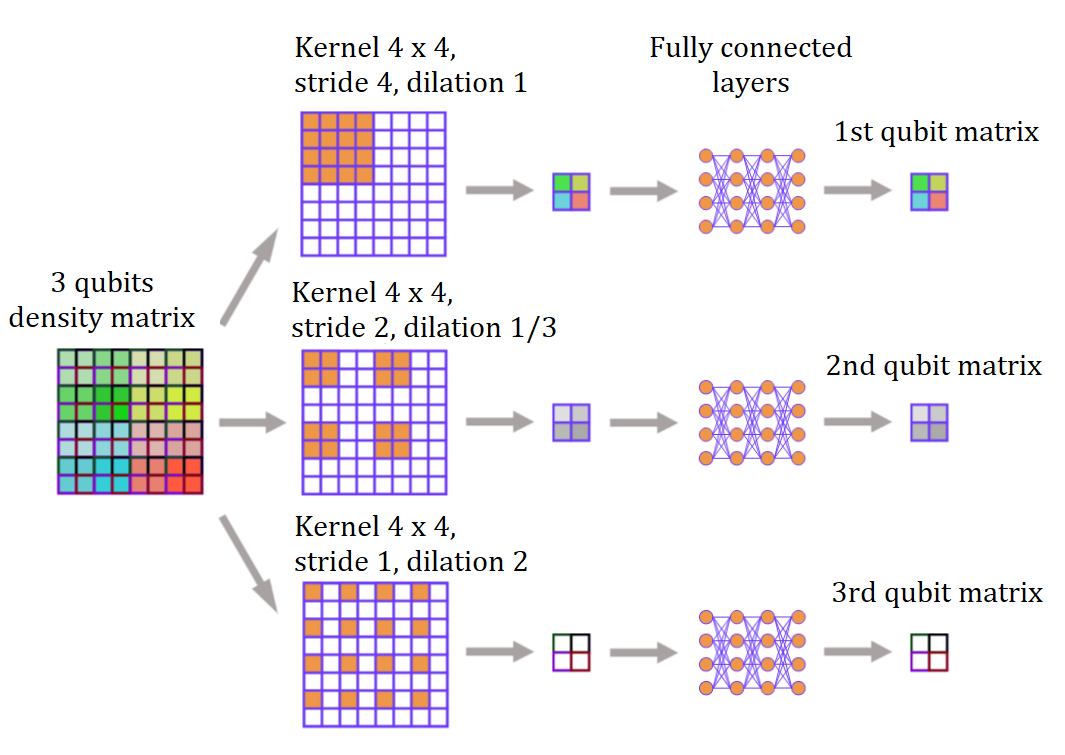}
    \caption{The 3-qubit neural network \textit{separator} model. The input density matrix is convolved separately with three convolution layers, each with $4\times4$ kernels that work in parallel.
    The \textit{stride} and \textit{dilation} parameters defining these convolution layers are arranged in such a way that they operate on each qubit subspace separately (details can be found in SM \cite{sm}). After the convolution layers, shape-preserving fully-connected layers are applied independently for each qubit matrix.}
\label{fig:separator}
\end{figure}

 To build the model, we use convolutional NNs (with a discrete convolution operation~\cite{zeiler2014visualizing,Yu2016dilated,dumoulin2016guide})
 as they are commonly applied to build robust representations of correlations on data on regular grids such as matrices. First, we need to organize the convolution layers, so that the resulting NN can separate the input density matrix into single qubits matrices. The constructed NN should preserve pure separable (product) states by definition and also not change the separability of mixed states. A closer look at the model scheme presented in Fig.~\ref{fig:separator} shows that the convolution layers are organized in a way that is a generalization of the partial trace operation, in the sense that the kernel simplified to identity matrix would calculate the partial trace over the remaining qubits. 
Such an architecture implies qubit permutation equivariance, meaning that permutation of qubits in the input matrix gives the same permutation of the reconstructed state and preserves separability of any reconstructed state.
 
 Additionally, 
 to extend the model complexity 
 we include four fully-connected (FC) layers with nonlinear activation functions (multilayer perceptron) after the convolutional ones. 
 This extension has a physical justification that will become clear later.
 Importantly, to preserve the separability property, these layers are applied independently to each block representing the qubit matrix.
 Moreover, to allow the neural network to generalize and reconstruct also separable mixed states, more sets of convolutions, i.e. \textit{channels}, have been used, giving a separable density matrix that is a sum over $N_K$ channels.
 Details can be found in the Supplemental Material (SM) \cite{sm}.  
 The reconstructed density matrix is given by
\begin{equation}
    \hat{\rho}_{ABC} = \frac{1}{N} \sum_{i=1}^{N_K} \hat{\rho}_{Ai} \otimes \hat{\rho}_{Bi} \otimes \hat{\rho}_{Ci},
    \label{eq:Kronecker}
\end{equation}
where we calculate the sum of the Kronecker product of output single qubit matrices ($\hat{\rho}_{Ai}$, $\hat{\rho}_{Bi}$, $\hat{\rho}_{Ci}$) over different $N_k$ channels, numbered by $i$. Note that the reconstructed matrix is separable, but, in general, it does not fulfill the additional criteria which would
make it a non-discordant state from definition.
Since a perfect decoupling into single-qubit matrices is possible only for separable states, the minimal distance between the original and reconstructed density matrices, obtained by the training of the neural network, can be seen as a measure of the separability of states.

Both the original and reconstructed density matrices have to be included in the NN loss function, 
which describes how different the reconstructed state is from the original state.
In quantum information theory, one of the natural choices for a difference between density matrices is the Bures distance \cite{laha2021}, but numerical experiments showed much better efficiency for training our NN model with the well-established $L^1$ norm, 
\begin{equation}
    \mathcal{L}(\rho_{ABC}) = \frac{1}{d^2} \sum_{i,j} |\rho^{ij}_{ABC} - \hat{\rho}^{ij}_{ABC}|.
    \label{eq:loss}
\end{equation}
Here $\rho^{ij}_{ABC}$ and $\hat{\rho}^{ij}_{ABC}$ denote the elements of the original and reconstructed density
matrices labeled by the indices $i$ and $j$, and
$d^2$ is the number of matrix elements. 
Using this loss function, we train the NN model
so that the total reconstruction loss, summed over all training examples, is minimized. 
Note that to train the network, 
which we will call a \textit{separator},
we do not need the labeled data. Instead, we compare inputs with outputs, which means
that our training scheme is fully unsupervised.
These types of ML models are also called {\it autoencoders} (we comment on this in more detail in SM \cite{sm}). 

Importantly, by choice of the model architecture, the separator network
is suited to reconstruct separable states,
hence, for these states we expect the value of the loss function to be close to zero.
In contrast, for entangled states, the model is not able to reconstruct the input, and the difference between the two matrices
will result in a significantly higher value of the loss $\mathcal{L}$. In order to distinguish between separable and entangled states, we set the anomaly threshold $\tau$ in such a way that if $\mathcal{L}(\rho_{ABC})$ is greater than $\tau$ for a given state $\rho_{ABC}$, then this state is classified as entangled. Otherwise, it is marked as separable. 

The most critical aspect of training an ML model is having a well-balanced and diversified data set. The modern approach in ML is to synthesize data with already known properties (e.g.~forming separate classes in case of classifier training). To achieve this goal, we select different methods for generating random density matrices. The main idea is to generate random states using the quantum circuits approach \cite{siamese2022}, but in order to have a well-diversified data set, we also use a technique based on sampling from the uniform Haar measure~\cite{mezzadri2006} and an additional parameterized method designed by us (generation details can be found in SM \cite{sm}). 

We construct a
training set composed of 530~000 pure and mixed separable states including a class of non-discordant states
and a validation set of 50~000 is constructed in the same way.
Furthermore, we generate two test sets involving both separable and entangled states. The first of those ($S_\mathrm{pure}$) is constructed from 15~000 random pure separable states and 15~000 random pure entangled states. This data set is generated evenly using both the quantum circuit approach and sampling from the Haar measure. The second test set ($S_\mathrm{mixed}$) includes 65~000 mixed states, constructed from pure states either by mixing different states with random probabilities or by reducing a larger multi-qubit space to the desired few-qubit size via the partial trace.

To classify states as separable during the generation process, we use the \textit{negativity} \cite{vidal02,plenio05b}, an entanglement measure
based on the PPT-criterion of entanglement \cite{peres96a,horodecki96},  which is one of the few measures that can be
found directly from the density matrix for systems larger than two qubits. It is defined as the absolute sum
of the negative eigenvalues of a bipartite density matrix after partial transposition is performed
on either of the subsystems. 
The generation methods for mixed entangled states are specifically adjusted not to produce bound entangled states \cite{horodecki99},
which cannot be detected by negativity.
To qualify discord, we use the criterion of Ref.~\cite{huang11}, which is very straightforward to check,
especially if one of the subsystems is small (relevant details are given in SM \cite{sm}). Incidentally, it is common that a state is discordant only with
respect to one part of a bipartition, so a state has to contain no discord with respect to both subsystems to be classified as a non-discordant state \cite{modi14}.
For both entanglement and discord, the lack of correlations is checked three times: between qubit $i$ and the rest of the 3-qubit system,
with $i=A,B,C$.

\begin{table}[b]
\begin{tabular}{|l|cccc|}
\hline
\multicolumn{1}{|c|}{
\begin{tabular}[c]{@{}c@{}}test\\ set\end{tabular}
}
& \multicolumn{4}{c|}{$\langle\mathcal{L}\rangle$ for different subsets} \\ \cline{2-5} 
\multicolumn{1}{|c|}{} & \multicolumn{1}{c|}{separable} & \multicolumn{1}{c|}{non-discordant} & \multicolumn{1}{c|}{discordant} & entangled \\ \hline
$S_\mathrm{pure}$ & \multicolumn{1}{c|}{$\num{5.9e-4}$} & \multicolumn{1}{c|}{$\num{5.9e-4}$} & \multicolumn{1}{c|}{$0.042$} & $0.042$ \\ \hline
$S_\mathrm{mixed}$ & \multicolumn{1}{c|}{$\num{5.2e-3}$} & \multicolumn{1}{c|}{$\num{1.0e-3}$} & \multicolumn{1}{c|}{$\num{0.017}$} & $\num{0.021}$ \\ \hline
\end{tabular}
\caption{Average separator loss for different subsets in pure $S_\mathrm{pure}$ and mixed $S_\mathrm{pure}$ test sets quantifying similarity between different types of states and separable ones.}
\label{tab:loss}
\end{table}

In ML a single pass through the entire training set, during which we update model parameters (typically using the gradient of a loss function in the parameter space), is called an epoch. The separator model was trained for 20 epochs, in a way that the model parameters for the epoch with the lowest averaged reconstruction loss $\langle\mathcal{L}\rangle$ on the validation set were saved and used for evaluation (testing) purposes. 

For the test set that contains only pure states, $S_\mathrm{pure}$, the average loss $\langle\mathcal{L}\rangle$ for separable states is equal to $\num{5.9e-4}$, while in the case of entangled states, it is much larger, giving $\langle\mathcal{L}\rangle=0.042$. Obviously, for the discord the numbers are the same -- see Table~\ref{tab:loss}. Choosing the threshold value $\tau$ to obtain the highest possible accuracy (balanced accuracy, $BA$, defined in SM \cite{sm}) of the model predictions yields $\tau=\num{1.25e-3}$ with the corresponding classification accuracy of $\num{99.63}\%$ for the whole $S_\mathrm{pure}$ set. 

For the test set with mixed states, $S_\mathrm{mixed}$, in the case of non-discordant states, the average loss $\langle\mathcal{L}\rangle$ is only about $1.5$ times higher than for uncorrelated pure states,
meaning that the original and reconstructed states are the same to an equivalent extent, while for mixed separable
states it is $40$ times higher, signaling that, on average, mixed separable states differ significantly more from the reconstruction.
Moreover, discordant and entangled states display roughly half $\langle\mathcal{L}\rangle$ for mixed states
compared to pure correlated states, which is in agreement with the fact that mixed states are on average less
correlated than pure states. 
Overall, the difference between
zero and nonzero discord states (measured by reconstruction loss $\mathcal{L}$) is over an order of magnitude larger than between separable and entangled states.

Let us now test the separator performance using the mixed test set, $S_\mathrm{mixed}$, and compare its practical usefulness for either discord or entanglement detection.
We performed two independent tests. In the first, we took discord labels and tested the ability to classify discord, and in the second, we took separability labels for the same set $S_\mathrm{mixed}$ and tested the ability to classify entanglement. 
Since for mixed states, a discordant state can also be separable, one cannot simultaneously balance non-discordant vs.~discordant and separable vs.~entangled states in a single mixed test set, therefore we use $BA$ (see SM \cite{sm}) as the model performance measure.
\begin{figure}[!tb]
    \centering \includegraphics[width=\columnwidth]{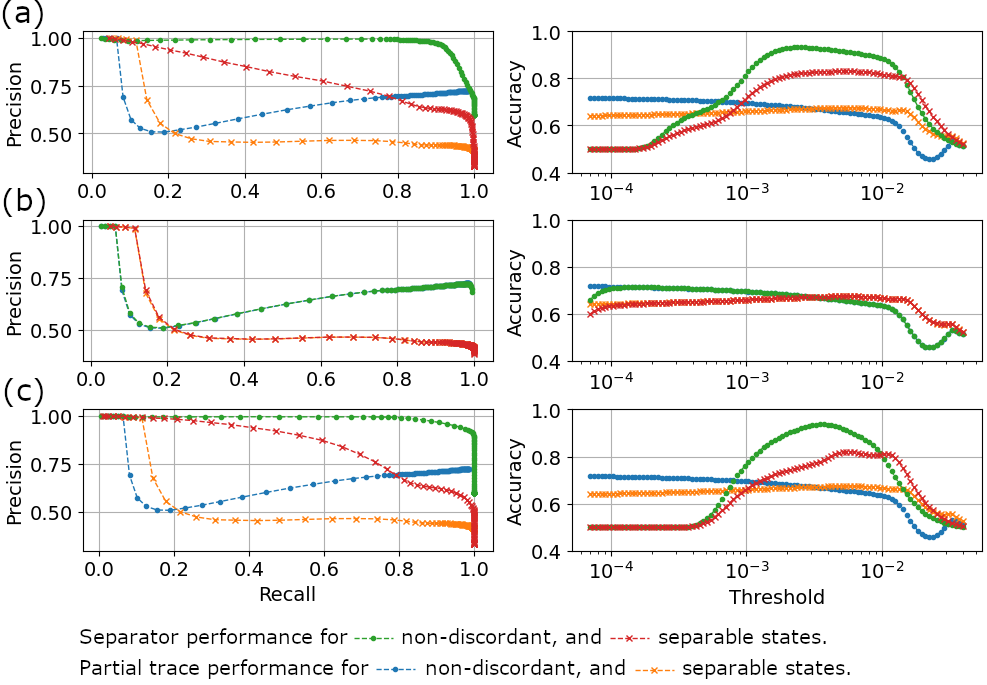}
    \caption{(a) Separator performance for detecting discordant (green) and entangled (red) states, tested on $S_\mathrm{mixed}$. The results present: (left) precision vs. recall curves, and (right) balanced accuracy, $BA$ depending on the threshold $\tau$ value. For comparison, partial trace-based baseline model performance is also presented by blue and orange curves, respectively. (b) Same as (a) but for separator model with removed FC layers -- cf.~Fig.~\ref{fig:separator}. (c) Same as in (a) but for the separator trained on the non-product states only.}
    \label{fig:similarity_plot_0}
\end{figure}

Fig.~\ref{fig:similarity_plot_0}
contains results of separator performance for detecting discordant (green curves) and entangled (red) states on the $S_\mathrm{mixed}$ set.
Fig.~\ref{fig:similarity_plot_0}(a), right, shows that the model has much higher $BA$ when detecting discord than entanglement in a wide range of thresholds, reaching $BA=93$\% for $\tau=\num{0.0023}$ in case of discord detection, while for entanglement the
maximum $BA=82$\% is reached at $\tau=\num{0.0051}$.
Moreover, when training the model on the subspace of the original training domain where the product states have been excluded, as presented in Fig.~\ref{fig:similarity_plot_0}(c), we clearly get better performance for discord in terms of the larger area under the precision-recall curve (left). Performance comparisons for training separately on some other classes of states can be found in SM \cite{sm}.

The convolution part in the separator performs an operation that is a generalization of the partial trace. When removing the FC layers that extend the convolution layers, we observe that the model simply collapses to the partial trace baseline (example kernels for both cases can be found in SM \cite{sm}) and is analogous to what is obtained when training only on pure separable states. Fig.~\ref{fig:similarity_plot_0}(b) contains results for a
separator with the FC layers removed: comparing accuracy with a partial trace baseline (blue and orange curves) we get practically the same accuracies.

We present the confusion matrices for the separator serving as a separability/entanglement classifier in Table~\ref{tab:confusion_matrix_entanglement} and a non-discordance/discordance classifier in Table~\ref{tab:confusion_matrix_discord}.
Overall, the separator tends to grossly overestimate the set of entangled states as seen in Table~\ref{tab:confusion_matrix_entanglement}, where for both thresholds presented ($\tau=0.0023$ corresponding to the best
discord classification and $\tau=0.0051$ to the best entanglement classification) the number of separable states
wrongly labeled as entangled is of the same order of magnitude as the number of correctly detected entangled states.
The number of erroneously classified separable states is comparatively small. 
On the contrary, the set of discordant states tends to be underestimated, with a discordant state being more likely to be labeled as
not correlated than the other way around. This occurs to a much lesser extent than the overestimation of entanglement as seen in Table~\ref{tab:confusion_matrix_discord}. The results here are more balanced, with erroneous classifications averaging
at a number an order of magnitude lower than the correct predictions. 
\begin{table}[b]
\begin{tabular}{|l||*{2}{c|}}\hline
\backslashbox{Labels}{Predictions}
&\makebox[7em]{separable}&\makebox[7em]{entangled}\\\hline\hline
separable $\hspace{1.2em}\tau=0.0023$ & 28340 & 15257\\\hline
entangled $\hspace{1em}\tau=0.0023$& 305 & 21098 \\\hline\hline
separable $\hspace{1.2em}\tau=0.0051$& 30696 & 12901 \\\hline
entangled $\hspace{1em}\tau=0.0051$& 942 & 20461 \\\hline
\end{tabular}
\caption{Confusion matrix for separable/entangled states classification at $\tau = 0.0023$
and $\tau = 0.0051$.}
\label{tab:confusion_matrix_entanglement}
\end{table}
\begin{table}[t]
\begin{tabular}{|l||*{2}{c|}}\hline
\backslashbox{Labels}{Predictions}
&\makebox[7em]{non-discordant}&\makebox[7em]{discordant}\\\hline\hline
non-discordant $\tau=0.0023$& 25114 & 1192 \\\hline
discordant $\hspace{0.7em}\tau=0.0023$& 3531 & 35163 \\\hline\hline
non-discordant $\tau=0.0051$ & 25822 & 484 \\\hline
discordant $\hspace{0.7em}\tau=0.0051$& 5816 & 32878 \\\hline
\end{tabular}
\caption{Confusion matrix for non-discordant/discordant states classification at $\tau = 0.0023$
and $\tau = 0.0051$.}
\label{tab:confusion_matrix_discord}
\end{table}

\begin{figure}[!tb]
\centering
\includegraphics[width=0.9\columnwidth]{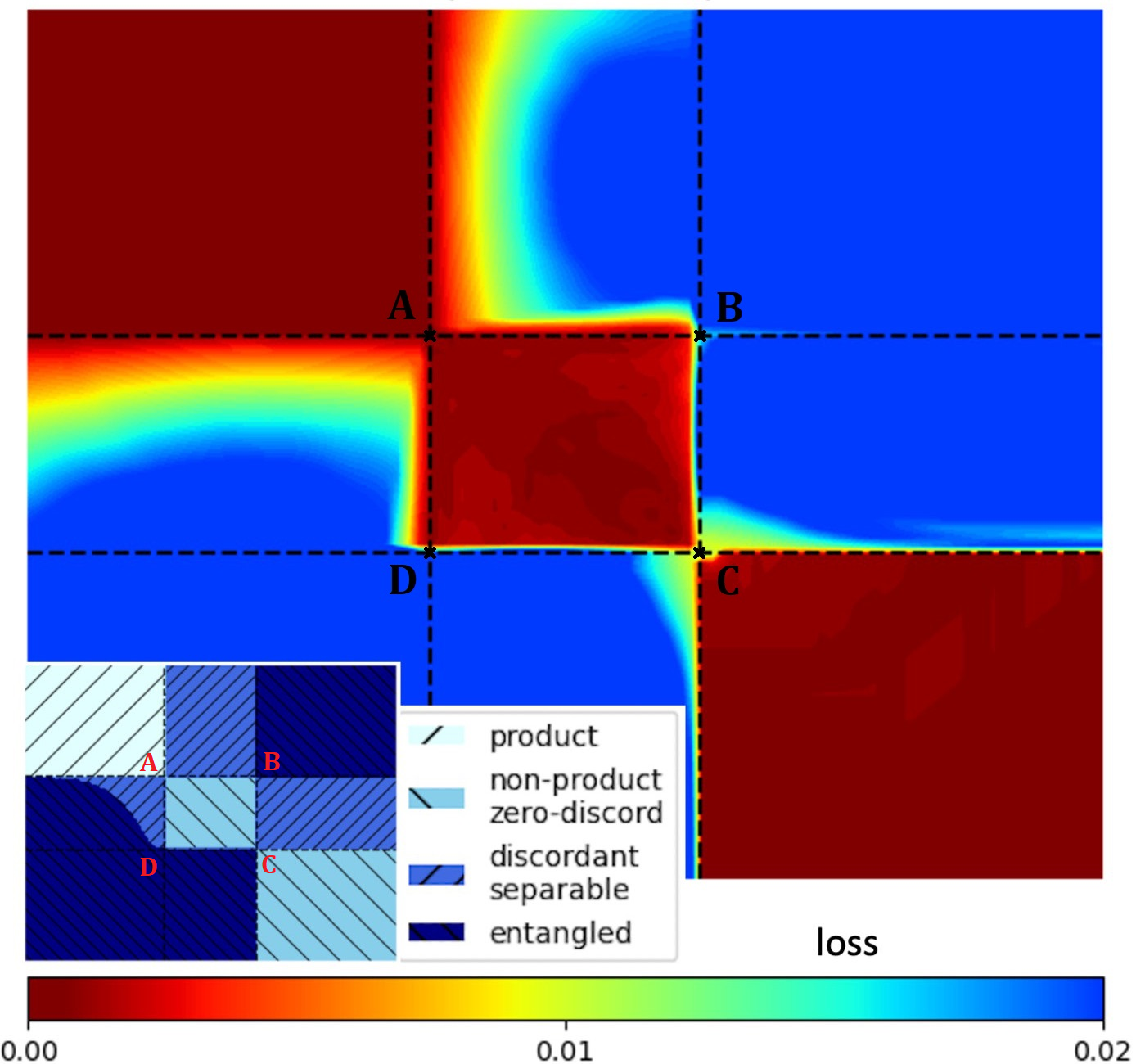} \\
  \caption{Reconstruction loss $\mathcal{L}$ for the already trained separator model tested on the family of 3-qubit states with their separability/discordance conditions known and parameterized on 2D map (see inset for the map division into the classes).}
  \label{fig:maps}
\end{figure}

In order to acquire a deeper understanding of how well the sets of non-discordant and separable states are reconstructed by the ML classifier at different thresholds,
we analyze a family of 3-qubit states which are easy to parameterize, and study a 2D diagram forming a map with regions differing in the type of correlations present (or absent)
as seen in the inset of Fig.~\ref{fig:maps}.

The family of states is constructed as follows. We take a separable mixed state of the form $\rho = p\rho_1 + (1-p)\rho_2$,
where probability $p$ is one of the parameters, and the density matrices $\rho_1$ and $\rho_2$ correspond to 3-qubit
pure states, $|\Psi_i\rangle =\bigotimes_k|\psi_i\rangle$, with $k=A,B,C$ and 
\begin{subequations}
\begin{eqnarray}
    |\psi_1\rangle&=&a|0\rangle+\sqrt{1-a^2}|1\rangle,\\
    |\psi_2\rangle&=&e^{-i\frac{\phi}{2}}\sqrt{1-a^2}|0\rangle-ae^{i\frac{\phi}{2}}|1\rangle,
\end{eqnarray}
\end{subequations}
respectively. Coefficient $a$ and phase factor $\phi$ constitute two other parameters that characterize the family of states. It
now contains only separable states. For $p = 0, 1$ the state is pure (and consequently of product form), while it is mixed non-discordant for $\phi = 0$ or $a = 0, 1$.  
To introduce entanglement into the system we find the spectral decomposition of $\rho=\sum_i \lambda_i |\varphi_i\rangle \langle\varphi_i|$ and increase the largest eigenvalue, $\lambda_0$,
\begin{equation}
    \rho \rightarrow \frac{\sum_{i \neq 0} \lambda_i |\varphi_i\rangle \langle\varphi_i| + (\lambda_0 + c)|\varphi_0\rangle \langle\varphi_0|}{\sum_{i} \lambda_i + c},
\end{equation}
making $c$ the final parameter which characterizes this family of states.

This allows us to plot the loss function (\ref{eq:loss}) for selected cross sections of the Hilbert space and visualize
how surfaces of equal $\mathcal{L}$ compare to the sets of non-discordant and separable states.
In Fig.~\ref{fig:maps} only the darkest regions in the inset are entangled, all other regions are separable and are further divided
into product states, non-discordant mixed states, and discordant separable states.
The respective regions are obtained by careful choice of parameters $p$, $a$, $\phi$, and $c$ along the horizontal and vertical axes. 

The main plot in Fig.~\ref{fig:maps} shows the loss $\mathcal{L}$ on this parameter space. 
The specifics of how the different parameters change in each region are described in detail in SM \cite{sm}. It is noticeable that, while the partial trace-based model is capable only of distinguishing between product and non-product states
(we show the plot and discuss this in detail in SM \cite{sm}), the neural network distinguishes all non-discordant states.
The results presented in Fig. \ref{fig:maps} are in agreement with the general results of Fig.~{\ref{fig:similarity_plot_0}},
and Tables \ref{tab:confusion_matrix_entanglement} and \ref{tab:confusion_matrix_discord}, demonstrating 
why the discord detection results in false negatives (overestimating the set of non-discordant states), while entanglement detection results in so many false positives.

In conclusion, we designed and trained a neural network to distinguish between separable and
entangled states on a set of mixed and pure states. For optimum threshold values, the network
performs with $82\%$ balanced accuracy for mixed states, 
as compared to over $99\%$ classification accuracy of the same network for
pure state entanglement. To understand this discrepancy outside of the trivial acknowledgment
that determining mixed state entanglement is a much more complicated task than pure state
entanglement, we have compared the results obtained on the mixed test set with
other classes of uncorrelated states. We have found that the set of states labeled as uncorrelated
by the separator overlaps best with the set of non-discordant states. This is confirmed by the $93\%$ balanced accuracy of detecting such states, outperforming the detection of both separable and product states. This result is highly surprising, because it shows that, by ML determination, random mixed non-discordant states are much more similar to states from a random set of pure separable states (as the lowest loss is observed for such states) than random mixed separable states.
We supplement plots of ML metrics and confusion matrices with a map of numerically detected quantum correlations
for cross sections of the 3-qubit Hilbert space, which allows us to understand the nature of the sets of states which
are qualified as uncorrelated by ML.  The surface where the loss
is near zero (no quantum correlations) reproduces the set of non-discordant states extremely well,
while the discordant separable states are largely classified as quantum-correlated, illustrating
why the separator grossly overestimates entanglement while only marginally underestimating
discordance. 
The network architecture is designed carefully to preserve
separability (i.e. it does not introduce any entanglement when reconstructing the output) and its
output is equivariant with respect to qubit permutations. We observe that the simplified version of
the separator model, which realizes just the generalization of a partial trace operation, simply collapses
to partial trace during the training. However, when extending it to a neural model capable to learn
other representations by adding fully-connected layers working separately within each qubit subspace, it learns more complex patterns and improves its performance over a partial trace baseline.

\textit{Acknowledgements.} M.K. and J.P. acknowledge support from National Science Centre (Poland) under Grant No. 2021/43/D/ST3/01989.
M.M.M. acknowledges support from the National Science Centre (Poland) under Grant No. DEC-2018/29/B/ST3/01892.

\bibliography{bibliography}

\begin{thebibliography}{54}
\bibitem [{sm()}]{sm}%
  \BibitemOpen
  \bibfield  {title} {\bibinfo {title} {Url will be inserted by publisher},\
  }\href@noop {} {\ }\BibitemShut {NoStop}%
\end{thebibliography}

\clearpage
\setcounter{equation}{0}
\setcounter{figure}{0}
\setcounter{table}{0}
\makeatletter
\renewcommand{\theequation}{S\arabic{equation}}
\renewcommand{\thefigure}{S\arabic{figure}}
\renewcommand{\bibnumfmt}[1]{[S#1]}
\clearpage
\onecolumngrid
\begin{center}
    {\large\bf Supplemental Material for ``Data-driven criteria for quantum correlations''}
\end{center}
\setcounter{page}{1}
\vspace*{4mm}

\twocolumngrid
\subsection{Details of the  separator neural network}
In the following section, we analyze the structure of the neural network (NN) model trained to reconstruct separable states and detect non-separable states as an anomaly. 
\begin{figure}[b]
    \centering
\includegraphics[width=0.32\textwidth]{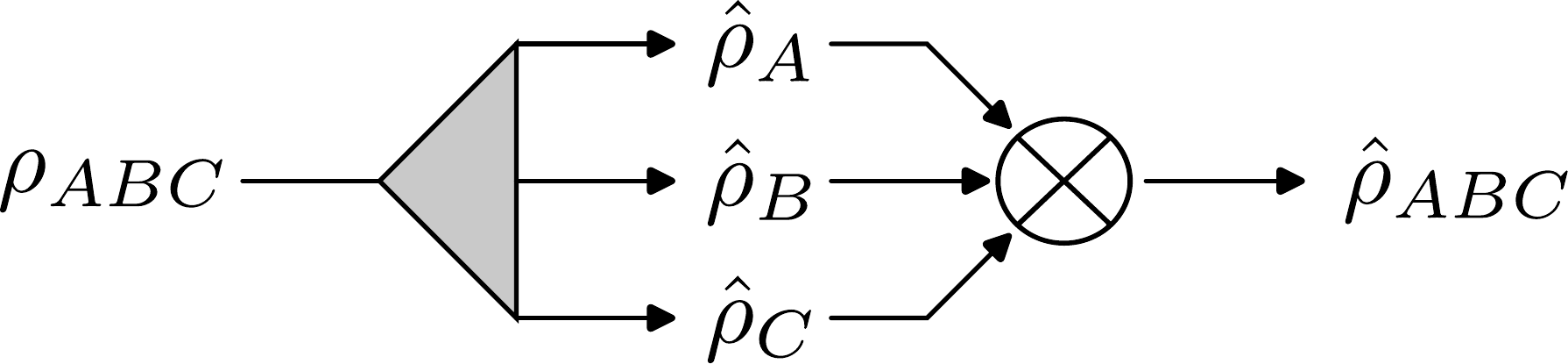}
    \caption{The separator autoencoder composed of trainable encoder neural network (gray triangle) and analytical decoder that simply calculates the Kronecker product. }
    \label{fig:splitter}
\end{figure}
It is easy to construct a many-qubit density matrix as a product of single-qubit density matrices. However, the opposite operation is much more complicated. Then, this task can be seen as an {\it inverse problem}. Here, we propose to solve it with the help of a modified autoencoder-type NN, schematically depicted in Fig.~\ref{fig:splitter}. The typical autoencoder is composed of two connected networks: the {\it encoder} and the {\it decoder}. Usually, the output of the encoder is narrower than its input, so this part compresses the data. Next, the decoder decompresses it, and both parts of the autoencoder are trained to have the output as close as possible to the input. Our idea is to use the encoder to transform 3-qubit density matrices into three single-qubit matrices in such a way that their product, calculated by the decoder, is as close to the original density matrix as possible. The autoencoder will be implemented as a NN, with only the encoder part trainable. The decoder simply calculates the Kronecker product of the single-qubit density matrices, as described in Eq.~(\ref{eq:Kronecker}). 

The backbone of the encoder is composed of convolution layers, which map the input density matrix into single qubit subspaces. They are followed by fully connected (FC) layers that extend the model and introduce non-linearity; however, they do not introduce entanglement, since they are applied independently for each of the reconstructed qubits.

Let us focus on a $d=3$ qubit system $\rho_{ABC}$, bearing in mind that the proposed reconstruction can be easily extended to larger qubit registers. We require a neural network model $\mathcal{N}$, which we call a separator, to have the following properties:
\begin{align}
     &\mathcal{N}(\rho_A \otimes \rho_B \otimes \rho_C) = c\,\rho_A \otimes \rho_B \otimes \rho_C,\label{eq:nn1}\\
     &\mathcal{N}\left(\sum_i p_i\,\rho_{Ai} \otimes \rho_{Bi} \otimes \rho_{Ci}\right) = \sum_i \hat{p}_i\,\hat{\rho}_{Ai} \otimes \hat{\rho}_{Bi} \otimes \hat{\rho}_{Ci},\label{eq:nn2}
 \end{align}
 with some multiplicative constant $c$.
To do so one can adjust the convolution layers $\mathcal{C}^{s,l}$, with kernel $\mathcal{K}$, step size (stride) $s$, and dilation $l$:
\begin{equation}
\mathcal{C}_{s,l}(\rho)^{i,j}=(\mathcal{K}\ast_{l}\rho)^{si,sj}=\sum_{k,q}\mathcal{K}^{k,q}\rho^{si+l(k),sj+l(q)},
\end{equation}
$i,j=0,1$, and $k,q=0,...,2^{d-1}-1$,
so that the kernels overlap with the respective blocks of the density matrix and therefore convolving them against the density matrix results with output matrices representing systems with a reduced number of qubits. 
Dilation can be just a simple scaling of the sum index: $l(k)=l\,k$, with rate $l$ meaning that the kernel only touches the signal at every $l$-th entry, or more complex indexing.

Specifically, choosing $\mathcal{C}_A=\mathcal{C}_{4,1}$ with kernel $\mathcal{K}$ of size $4\times4$ allows the neural network to learn the structure of repeating blocks $\rho_B \otimes \rho_C$ and therefore to reduce the input density matrix to: 
\begin{equation*}
\mathcal{C}_{4,1}(\rho_A \otimes \rho_B \otimes \rho_C)=c\,\rho_A,
\end{equation*}
with $c$ being some multiplicative constant, as in Eq.~(\ref{eq:nn1}).
To extract the last one, we need $\mathcal{C}_C=\mathcal{C}_{1,2}$.
To extract the second qubit, more complex dilation is needed in $\mathcal{C}_B$.
All kernels are visualized in Fig.~\ref{fig:separator}. 
The input density matrix, 
    divided into 2 channels (respectively for the real and imaginary parts), is convolved separately with 3 convolution layers, each with $4 \times 4$ kernels.
The first one is set up using a stride equal to 4, as a result producing a matrix corresponding to the first qubit subspace. In the second convolution, the kernel is divided into 4 blocks of size $2 \times 2$, separated with dilations resulting in the second qubit subspace. The third convolution consists of the kernel dilated with extra space between every weight, therefore extracting the information about the last qubit.
Finally, after applying all the convolutions, one can combine them and construct the output separable density matrix $\mathcal{C}_A(\rho_{ABC})\otimes\mathcal{C}_B(\rho_{ABC})\otimes\mathcal{C}_C(\rho_{ABC})=\hat{\rho}_{A} \otimes \hat{\rho}_{B} \otimes \hat{\rho}_{C}$.

It is obvious that in this way the defined separator model $\mathcal{N}$ fulfils Eq.~\ref{eq:nn1}. However, to enable the neural network to generalize and reconstruct separable mixed states, i.e., obey Eq.~\ref{eq:nn2}, a larger number of (triples of) kernels is chosen (we use $N_K=24$ kernels in each type of convolution layer).
To simplify the implementation and keep neural network values real, we doubled the number of kernels, that is, $2N_K$, so half of the channels processed the real and the other half the imaginary part of the density matrix.
At this point, we also add four FC layers for each qubit ($A,B,C$) separately to extend the complexity of the model.
Subsequently, we calculate the sum of the Kronecker product of output single qubit matrices ($\hat{\rho}_{Ai}$, $\hat{\rho}_{Bi}$, $\hat{\rho}_{Ci}$) over different kernels numbered by $i$ and arrive at the reconstructed density matrix as in Eq.~\ref{eq:Kronecker}.

Both the original and reconstructed density matrices are included in the loss function of the neural network defined in Eq.~\ref{eq:loss}.
Using this loss function, we train the separator model so that the total reconstruction loss, summed over all training examples, is minimized. 
Note that to train the network, we do not need the data to be labeled. Instead, we simply compare inputs with outputs, and therefore, our model scheme is fully unsupervised. The trainable weights of the kernels are adjusted using the procedure called gradient descent, where the gradients of the loss are calculated with respect to the weights. Subsequently, all parameters of the model, e.g. each kernel weights $\mathcal{K}_{ij}$ are updated as follows:
\begin{equation}
    \mathcal{K}^{ij}_\mathrm{new} = \mathcal{K}^{ij}_\mathrm{old} - \alpha \frac{\partial \mathcal{L}}{\partial \mathcal{K}^{ij}_\mathrm{old}},
\end{equation}
where $\alpha$ is the learning rate. 
We proceed similarly for weights in FC layers.
This procedure is repeated until the reconstruction loss (Eq.~\ref{eq:loss}) reaches (we hope) a global minimum, which means that the model is already trained. Importantly, by the choice of the model architecture, the separator network
is suited to reconstruct separable states,
and for these states, we expect the value of the loss function to be close to zero.
In contrast, in the case of an entangled state, the model will not be able to reconstruct the input, and the output density matrix should differ from the input one, resulting in a significantly higher value of the loss. Therefore, to distinguish between separable and entangled states, we set the anomaly threshold $\tau$ in such a way that if $\mathcal{L}(\rho_{ABC})$ is larger than $\tau$ for a given state $\rho_{ABC}$, then this state is classified as entangled and, otherwise, it is marked as separable. 
Interestingly, in addition to classification purposes, the loss value $\mathcal{L}(\rho_{ABC})$ can also be interpreted as a similarity measure between the states used in the training dataset and given tested state $\rho_{ABC}$. 

\begin{figure*}[!htb]
    \centering
    \includegraphics[width=1.4\columnwidth]{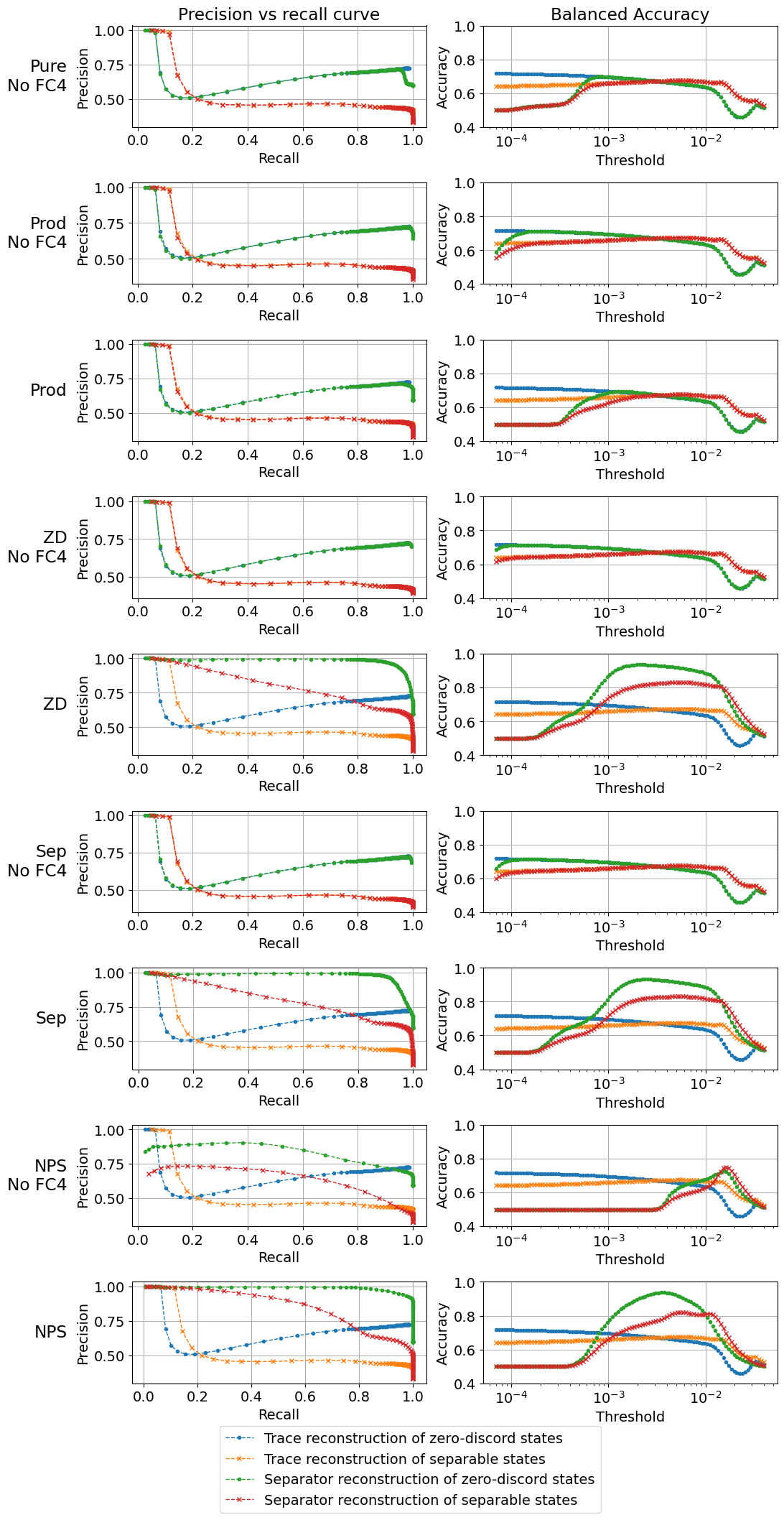}
    \caption{
    Same as in Fig.~\ref{fig:similarity_plot_0} but for separator trained only on some part of the original training set: pure separable states (Pure), product states (Prod), non-discordant states (ZD), separable states (Sep) and non-product separable states (NPS). Results for separator without four fully-connected layers (containing only convolutional ones) are marked as "No FC4".}
    \label{fig:similarity_plot}
\end{figure*}
In Fig.~\ref{fig:similarity_plot} we present the extended results for the separator model with and without four FC layers trained on different subsets of the original training set. As before, the neural network is evaluated on the $S_\mathrm{mixed}$ set. We start to train the model in the simplest possible scenario where dataset consists of 190000 pure separable states. The model trained on such a dataset collapses to the baseline trace model. It is expected behavior since in this case the baseline model is a sufficient criterion for distinguishing between entangled and separable states. The same situation occurs when we extend our training set to the wider set of product states by adding 120000 mixed product states. Interestingly, in this case, the neural network reflects partial trace operation even more precisely, which is revealed by a wider range of thresholds producing overlapping accuracy. Including FC layers, and thus introducing the nonlinearity, slightly limits this phenomenon, however, since the baseline model is still a precise criterion of separability in the case of product states, hence the neural network is not penalized for not learning more complex operations. The situation changes when we fatherly append 100000 non-product non-discordant states. Now, the baseline is no longer reliable and the network needs to learn extra features in order to correctly separate and reconstruct such states. Notably, the separator model is capable of achieving this task only when extra FC layers are present. Otherwise, it collapses to the baseline. This situation reoccurs for full training set consisting of 530000 separable states. For the sake of completeness, we additionally test our model after training it on the reduced domain of non-product separable states. The data set in this case is constructed separately from 360000 states, which are generated with the same methodology as for the original training set. The main difference of the model trained on such a subdomain is that even without any FC layers it does not collapse to a partial trace operation, which may suggest that using only non-product separable states for training can also be beneficial for the model with FC layers.  

\begin{figure}[t]
\centering
\includegraphics[width=\columnwidth]{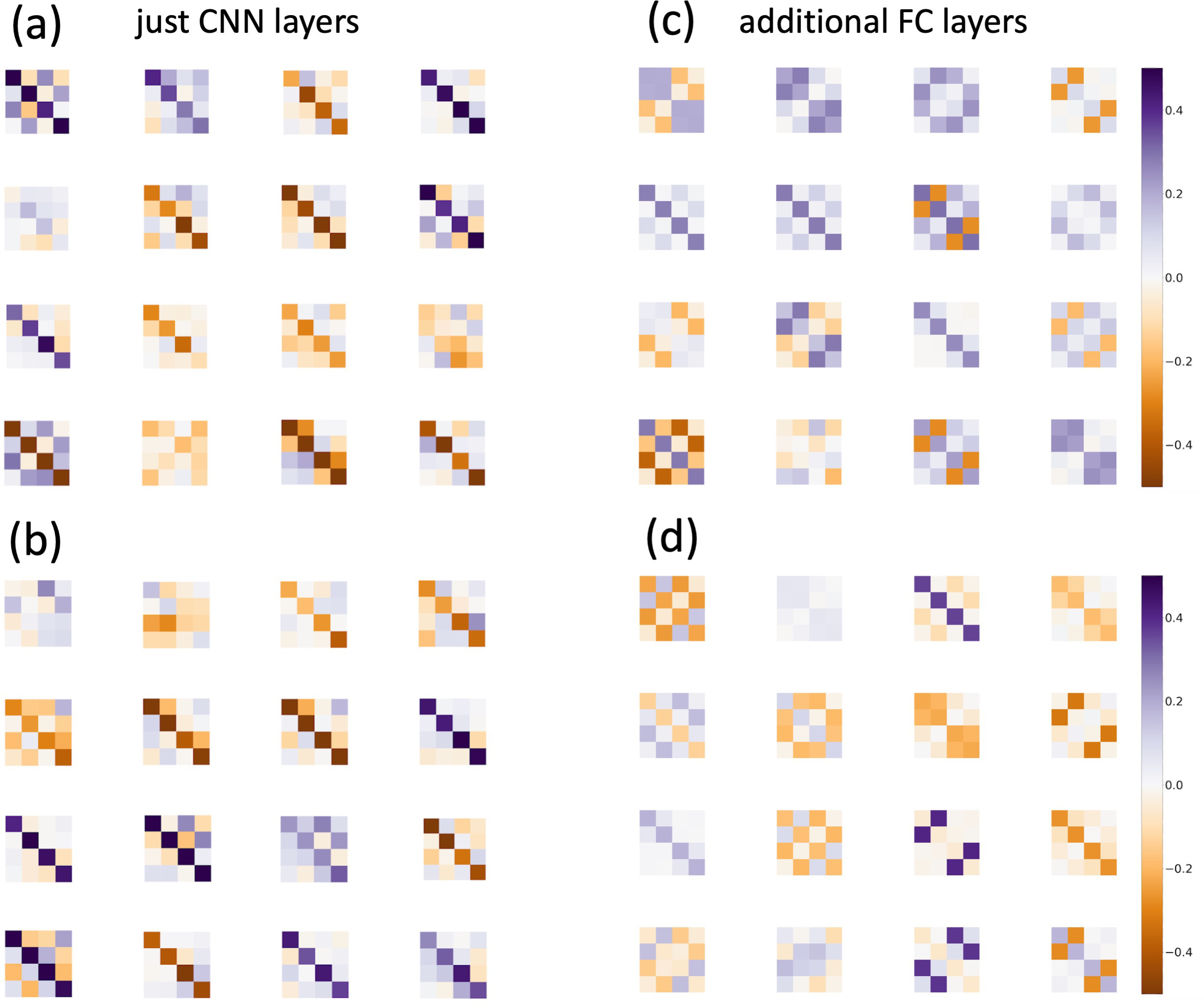}
\caption{Example weights of $4\times 4$ convolutional kernels for the already trained 3-qubit  separator model. The kernels reconstruct (a) real and (b) imaginary parts of the first qubit (weights for two other qubits are similar) for the plain separator model, with just convolutional layers. Similarly, (c) and (d) show real and imaginary parts of kernels for separator with additional fully-connected layers.}
\label{fig:kernels}
\end{figure}
In Fig.~\ref{fig:kernels} we present exemplary kernels learned for the separator model with (right) and without FC layers (left)
trained on the full data set (as in the main text). Model with only convolutional layers collapses to partial trace baseline which manifests itself in the fact that its kernels resemble identity matrices (partial trace is equivalent to the convolutional kernel being 
equal to the identity matrix).

\begin{figure}[b]
\centering
\includegraphics[width=0.90\columnwidth]{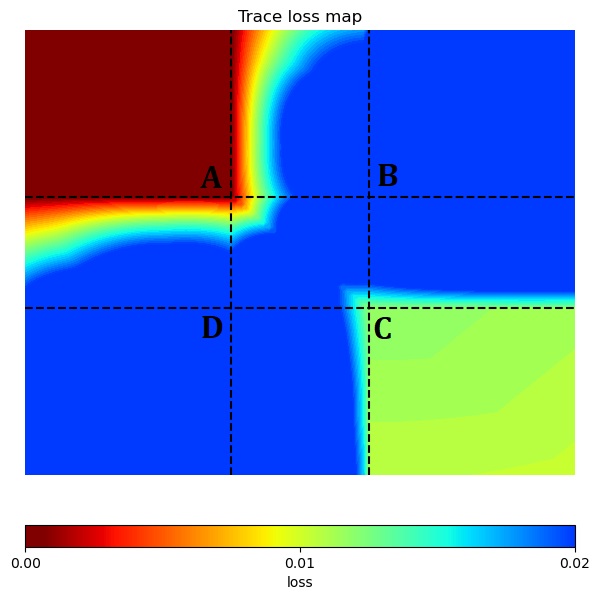}
  \caption{Reconstruction loss $\mathcal{L}$ for the partial trace-based model tested on family of 3-qubit states with their separability/discordance conditions known parameterized on 2D map.}
  \label{fig:maps_bm}
\end{figure}

In Fig.~\ref{fig:maps_bm} we also present a reconstruction loss map analogous to the map in Fig.~\ref{fig:maps}
in the main text, but obtained using a partial trace as a baseline replacement for the separator model.
On this map, only product states are clearly detected as not containing correlations, as is to
be expected; however, the region in the lower right square of the map, which represents some space of mixed zero-discord states,
is signified as containing fewer correlations. Note that the other region of similar type
(in the center of the plot) is not correspondingly distinguished. 

\subsection{Classification metrics}

In machine learning to evaluate classification or detection results, we typically use various types of metrics that measure model performance.

The standard metrics~\cite{Powers2011} used in binary classification, also to tune the model hyperparameters, e.g., classification threshold $\tau$, are precision $PR = \frac{TP}{TP + FP}$ and recall $RC = \frac{TP}{TP + FN}$.
Here $TP$, $TN$, $FP$, and $FN$ denote respectively the number of true/false positives/negatives and by negatives we mean non-discordant (or separable), while by positives -- discordant (or entangled) states. 

The procedure called precision-recall tradeoff involves choosing a threshold that maximizes the area under the precision-recall curve. However, due to the fact that our dataset is imbalanced (we cannot have balanced separable/entangled and non-discordant/discordant classes at the same time), we use a metric typically defined for imbalanced data sets. Balanced accuracy (BA) is defined as the average of recall obtained on each class \cite{balanced} as: $BA = \frac{1}{2}\left(\frac{TP}{TP + FN} + \frac{TN}{TN + FP}\right)$.


\subsection{Parameter changes on 3-qubit map}
The way in which the parameter space of the family of states depicted on the diagram map in Fig.~3.~is defined
is explained in the main text. Here, we detail how the specific parameters are changed
in the different regions of the map to generate states corresponding to different classes in terms of quantum and classical correlations.

Specifically, we start from point $A$, where the parameter $a = 1/{\sqrt{2}}$ takes the maximal value, while all others are set to minimal values: $p = 0$, $\phi = 0$ and $c = 0$. Moving along the axis $AB$ we increase the parameter $p$ so that it takes the maximal value $1/2$ in point $B$. Similarly, coefficient $a$ is decreased along the axis $BC$, so that in point $C$ it is equal to $0$. Moving outside the square region $ABCD$ we fix parameters $a$ and $p$ to the boundary values and start increasing the parameters $\phi$ and $c$. The phase $\phi$ grows linearly along both horizontal and vertical axes in such a way that inside the square $ABCD$ it is equal to $0$ and it is equal to $\pi/2$ at the edges of the map. The parameter $c$ is increased along the counter-diagonal in the top right square, taking the minimal value $0$ in point $B$ and maximal value $1$ in the top right corner of the map. The parameterization is performed symmetrically on both sides of the diagonal of the map. Therefore, the axis $AB$ in the upper right triangle corresponds to the axis $AD$ in the lower left triangle. Similarly, $BC$ is reflected to $DC$. Keeping that in mind, we note that parameter $c$ is additionally increased by $\Delta c$ in the lower left triangle. This extra term grows linearly along the counter-diagonal from $\Delta c = 0$ in the center of square $ABCD$ to $\Delta c = 1$ in the bottom left corner of the map.

\subsection{Parameterized method of generating 3-qubit density matrices}
In this section we describe the original parameterized method used to generate random mixed states when building the training set. To generate such states, we also used the quantum circuits approach and
a technique based on sampling from the uniform Haar measure -- both are described in \cite{siamese2022}.

Let us start by defining a 3-qubit pure separable state,
\begin{equation}
\begin{split}
    |\psi_{SEP}\rangle &= |\psi_1\rangle \otimes |\psi_2\rangle \otimes |\psi_3\rangle \\  &= (a_1|0\rangle +b_1|1\rangle) \otimes (a_2|0\rangle +b_2|1\rangle) \otimes (a_3|0\rangle +b_3|1\rangle) \\
    &= a_1a_2a_3|000\rangle + a_1a_2b_3|001\rangle + a_1a_3b_2|010\rangle \\ &+ a_1b_2b_3|011\rangle + a_2a_3b_1|100\rangle + a_2b_1b_3|101\rangle \\ &+ a_3b_1b_2|110\rangle + b_1b_2b_3|111\rangle, 
\end{split}
\end{equation}
where the complex coefficients of each state fulfil,
\begin{equation}
\label{abcd}
|a_i|^2+|b_i|^2=1.
\end{equation}
We then introduce entanglement into the system for each pair of qubits
via a phases $\phi_{ij}$,
\begin{equation}
\begin{split}
    |\psi_{ENT}\rangle &= a_1a_2a_3|000\rangle + a_1a_2b_3|001\rangle + a_1a_3b_2|010\rangle \\ &+ a_1b_2b_3e^{i\phi_{23}}|011\rangle + a_2a_3b_1|100\rangle \\ 
    & + a_2b_1b_3e^{i\phi_{13}}|101\rangle + a_3b_1b_2e^{i\phi_{12}}|110\rangle \\
    & + b_1b_2b_3e^{i(\phi_{12} + \phi_{13} + \phi_{23})}|111\rangle.
\end{split}
\end{equation}
Finally, the decoherence coefficients $c_i$ are included for each qubit separately in order to obtain mixed states, mimicking the operation of single-qubit dephasing channels,
\begin{align}
    \rho_{i} &= |\psi_i\rangle\langle\psi_i| = \left(
	\begin{matrix}
		|a_i|^2 & a_i b_i^*\\
		a_i^* b_i & |b_i|^2\\
	\end{matrix}
	\right) \\ &\xrightarrow{\mathrm{decoherence}} \left(
	\begin{matrix}
		|a_i|^2 & c_i a_i b_i^*\\
		c_i^* a_i^* b_i & |b_i|^2\\
	\end{matrix}
	\right) = \tilde{\rho}_i.
\end{align}
Combining all of the above, one arrives at the following:
\begin{equation}
    \rho = |\psi_{ENT}\rangle\langle\psi_{ENT}| \xrightarrow{\mathrm{decoherence}} \tilde{\rho},
\end{equation}
where the density matrix $\tilde{\rho}$ is parameterized with independent coefficients $a_i$ and $c_i$ as well as phases $\phi_{ij}$. In order to obtain a diversified set of states, we propose to choose these parameters uniformly at random over the interval $[0, 1]$ for $a_i$ and $c_i$, and over the interval $[0, 2\pi]$ for phases $\phi_i$. Moreover, we randomize the states by evolving them with local single qubit gates:
\begin{equation}
    U(\theta, \phi, \lambda) = \left( \begin{array}{cc} \cos\frac{\theta}{2} & -e^{i\lambda}\sin\frac{\theta}{2}  \\ e^{i\phi}\sin\frac{\theta}{2} & e^{i(\phi + \lambda)}\cos\frac{\theta}{2}  \end{array} \right),
\end{equation}
where $\theta$, $\phi$ and $\lambda$ are some arbitrary Euler's angles. Hence, sampling the angles randomly over the interval $[0, 2\pi]$, independently for separate qubits, one can write the final randomized density matrix as:
\begin{equation}
\label{rho_param}
    \tilde{\rho}_{RAND} = (U_1 \otimes U_2 \otimes U_3) \tilde{\rho} (U_1 \otimes U_2 \otimes U_3)^\dagger. 
\end{equation}

\subsection{Classifying states as non-discordant}
To classify a state as discordant or not, we
use an if-and-only-if criterion for non-discordant states from Ref.~\cite{huang11}.
The criterion states that a bipartite state with subsystems of 
arbitrary dimensions $N$ and $M$ has zero quantum discord
with respect to the system of dimension $M$,
if and only if all blocks of the $(NM)\times(NM)$ density matrix, which are obtained
by partitioning the density matrix into $N^2$ square matrices
of dimension $M$
are normal matrices and
commute with each other.

The required partition is performed starting from the density matrix written in the form
\begin{equation}
\hat{\sigma}=\sum_{kq}\sum_{nm}P_{kq}^{nm}|k\rangle \langle q|\otimes
|n\rangle \langle m|,
\end{equation}
where the indices and corresponding states $k,q$
correspond to the system of dimension $N$, while the indices and states
$n, m$ correspond to the subsystem of dimension $M$. 
The partition into blocks requires leaving only the elements which correspond to matrix elements $|k\rangle\langle q|$
in the subspace of the system of dimension $N$,
\begin{equation}
\label{partition}
\hat{\sigma}_{kq}=\langle k|\hat{\sigma}| q\rangle.
\end{equation}
Each choice of $k,q$ yields a different matrix.

A normal matrix is one that commutes with its Hermitian conjugate 
(the matrix obtained through the partition does not need to be density matrices), thus fulfilling
\begin{equation}
[\hat{\sigma}_{kq},\hat{\sigma}^{\dagger}_{kq}]=0,
\end{equation}
whereas commutation obviously requires that for all $k,q$ and $k',q'$ one has
\begin{equation}
[\hat{\sigma}_{kq},\hat{\sigma}_{k'q'}]=0.
\end{equation}
Both criteria are fulfilled, if and only if the state has zero discord
with respect to the subsystem of dimension $M$.

It is relevant to keep in mind that a zero discord state must not have any correlations of this
type both with respect to the subsystem of size $M$ and the one of size $N$, so in the case of our three
qubit system and each of the three qubit/2-qubits partitions, will require two checks for zero discord,
one with respect to the qubit, and the other with respec to the 2-qubit subsystem. In total, this means
that checking for the presence of discord in a given 3-qubit state is performed six times due to the 
different partitions and the asymmetry of the discord. 

\subsection{Numerical Methods}
Data generation is performed with the usage of qiskit~\cite{Qiskit} library. All data sets can be generated in \mbox{$\sim8$ hours}. Neural network model (encoder and decoder part) is implemented using Pytorch~\cite{paszke2019pytorch} library with an automatic differentiation engine (autograd) used to track gradient of the model parameters. The training of the single model on the GPU Nvidia RTX 3070 takes \mbox{$\sim1$ day}.

\end{document}